\documentclass[pra,twocolumn,superscriptaddress,notitlepage]{revtex4-1}
\usepackage{graphicx,amsmath,amsfonts,amssymb,upgreek,txfonts,color}
\usepackage[colorlinks,linkcolor=blue,citecolor=blue,urlcolor=blue,breaklinks=true]{hyperref}

\definecolor{defgray}{gray}{0.5}

\begin{document}

\title{Optomechanical characterization of silicon nitride membrane arrays}

\author{Bhagya Nair}
\author{Andreas Naesby}
\author{Aur\'{e}lien Dantan}
\email[Corresponding author: ]{dantan@phys.au.dk}
\affiliation{Department of Physics and Astronomy, University of Aarhus, DK-8000 Aarhus C, Denmark}

\date{\today}

\begin{abstract}
We report on the optical and mechanical characterization of arrays of parallel micromechanical membranes. Pairs of high-tensile stress, 100 nm-thick silicon nitride membranes are assembled parallel with each other with separations ranging from 8.5 to 200 $\mu$m. Their optical properties are accurately determined using a combination of broadband and monochromatic illuminations and the lowest vibrational mode frequencies and mechanical quality factors are determined interferometrically. The results and techniques demonstrated are promising for investigations of collective phenomena in optomechanical arrays.
\end{abstract}

\maketitle

Optomechanical resonators exploiting the radiation pressure forces of electromagnetic fields on high-quality mechanical oscillators have a broad range of applications ranging from metrology and fundamental physics tests to information processing and sensing~\cite{Aspelmeyer2014}. Integrated arrays of micro-/nano-mechanical resonators are currently intensely investigated owning to the promising prospects they offer, among others, for enhanced sensing performances and multimode/quantum optomechanics~\cite{Bhattacharya2008multiple,Hartmann2008,Ludwig2012,Xuereb2012,Stannigel2012,Seok2012}, or for the exploration of many-body physics phenomena, such as synchronization and heat transfer at the nanoscale~\cite{Heinrich2011,Ludwig2013,Zhang2012,Bagheri2013,Xuereb2014,Xuereb2015,Chesi2014,Schmidt2015}.  Various collective optomechanics platforms have been proposed and a number of multimode optomechanics experiments are beginning to be implemented, e.g. with capacitive drums in the microwave regime~\cite{Massel2012,Ockeloen2016} or microspheres~\cite{Dong2012}, microtoroids~\cite{Zhang2012,Zhang2015,Gilsantos2016}, nanobeams~\cite{Bagheri2013}, optomechanical crystals~\cite{Fang2016} and cold atoms~\cite{Botter2013,Spethmann2016} in the optical regime.

High-tensile stress micromechanical membranes made of low-loss material like silicon nitride have in this respect demonstrated excellent optomechanical properties and integrability in high-finesse optical resonators~\cite{Thompson2008,Wilson2009,Karuza2012,Kemiktarak2012NJP,Purdy2013}, with recent experiments involving the optomechanical dynamics of multiple membrane modes~\cite{Kemiktarak2014,Shkarin2014,Lee2015,Xu2016,Noguchi2016,Nielsen2016}. In addition to these remarkable features, the high degree of uniformity displayed by batches of simultaneously fabricated membranes or the possibility to realize electro-optomechanical interfaces~\cite{Bagci2014,Andrews2014,Fink2016} render them interesting candidates for the exploration of collective phenomena with optomechanical arrays~\cite{Xuereb2012}. In particular, exploiting structural optical resonances in periodic arrays of parallel membranes has been proposed as a means to dramatically enhance the optomechanical coupling strength as well as to tailor long-range, collective phonon-phonon interactions~\cite{Xuereb2012,Xuereb2013,Xuereb2014,Tignone2014}. 

We report here on the first realization of such arrays by assembling pairs of commercial, high-tensile stress silicon nitride membranes parallel with each other with separations ranging from 8.5 to 200 $\mu$m. Their optical transmission spectrum in the visible-near-infrared region is measured under either broadband and monochromatic illumination and the characteristic parameters of the arrays -- membrane thickness, refractive index and separation -- are extracted from a fit of the spectra to a one-dimensional model for multilayered systems. We show that these parameters, in particular the intermembrane separation, can be accurately determined and observe peak transmissions of up to $\sim$ 99.7\% at wavelengths around 900 nm, currently limited by imperfect parallelism in the assembly process. The lowest vibrational mode frequencies and mechanical quality factors are subsequently determined by means of optical interferometry in a low-pressure environment.\\

\begin{figure}[htbp]
\centering
\includegraphics[width=0.9\linewidth]{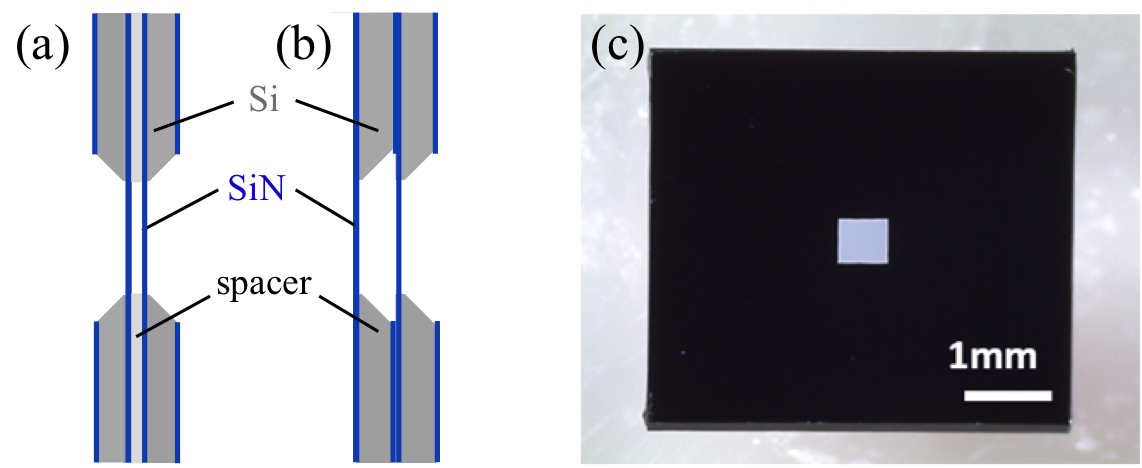}
\caption{Side view schematic of double-membrane array with (a) inbuilt spacer and (b) with Si frame spacer. (c) Top view photograph of array.}
\label{fig:fig1}
\end{figure}

\begin{figure}[t]
\centering
\includegraphics[width=\linewidth]{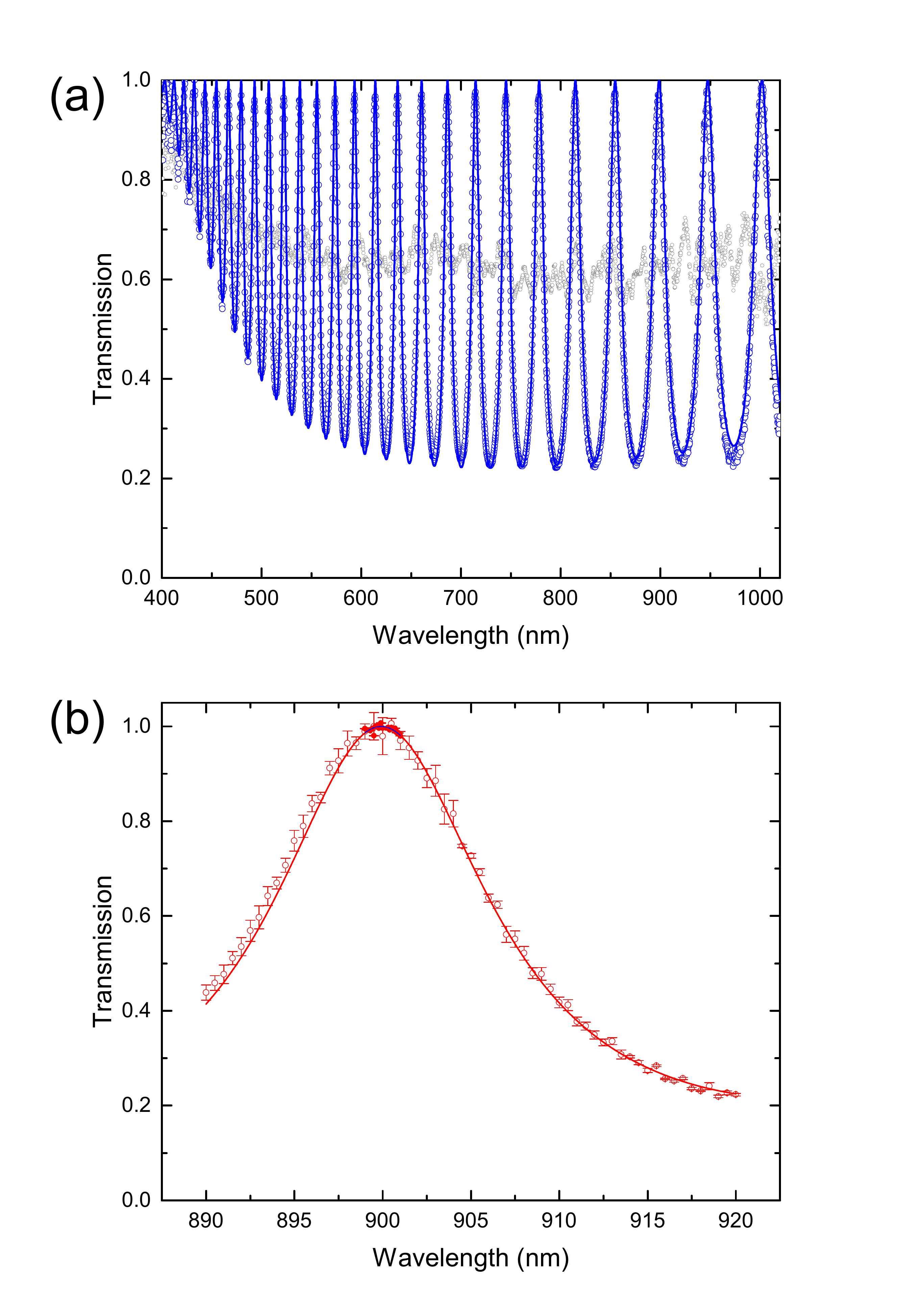}
\caption{(a) Transmission spectrum of a 8.5 $\mu$m-spacing array measured under (a) broadband and (b) monochromatic illumination. The solid lines represent the results of fits of the data with the theoretical model. The gray circles in (a) show the transmission of a single membrane as a reference.}
\label{fig:9um}
\end{figure}

The membranes used in this work are commercial (Norcada, Inc.), high-tensile stress ($\sim$ 1 GPa) square silicon nitride films with lateral dimension 0.5 mm and thickness 100 nm, deposited on a 5 mm-square silicon frame. As depicted in Fig.~\ref{fig:fig1}, pairs of membranes are assembled parallel with each other either {\it (i)} in a back-to-back configuration (Fig.~\ref{fig:fig1}a), in which a custom-made spacer (thickness $\sim$ 8.5 $\mu$m in this work) has been deposited on one of the chips, or {\it (ii)} on top of each other (Fig.~\ref{fig:fig1}b), using the silicon frame itself as a spacer (thickness 100 or 200 $\mu$m in this work). The membranes are positioned in a homemade holder and glued together on two frame sides with either epoxy or uv-curing glue. Prior to gluing the alignment of the membranes is performed in a simple fashion by illuminating the array with the focused radiation of a broadband visible source (flashlight tungsten filament), collecting the transmitted light with a multimode fiber and analyzing it with a spectrometer. An example of such a transmission spectrum, background-subtracted and normalized to that measured without sample, is shown in Fig.~\ref{fig:9um}a in the range 400-1020 nm for an array with an intermembrane separation of $\sim$8.5 $\mu$m. In contrast with the slowly varying transmission of a single membrane (Fig.~\ref{fig:9um}a, grey dots), the transmission spectrum of the array (blue dots) shows the expected interference pattern of a symmetric, low-finesse Fabry-Perot resonator, with near unity transmission at specific wavelengths.  

This spectrum can be adequately reproduced using a one-dimensional transfer matrix theory, in which the membranes are modelled as parallel dielectric slabs with thickness $l$ and refractive index $n(\lambda)$ separated by a distance $d$~\cite{Nair2016}. This model correctly accounts for the variation of the membrane reflectivity with the wavelength, as well as the phase-shift variation across the dielectrics, whose extension is not negligible with respect to the effective wavelength. Assuming the membranes to be identical and lossless and following Ref.~\cite{Nair2016}, it can be shown that the transmission of the array is given by
\begin{equation}
T(\lambda)=\frac{1}{1+m(\lambda)\cos(\phi-\psi)^2}\,,
\end{equation}
where $m(\lambda)=4\zeta(\lambda)[1+\zeta(\lambda)]^2$ is the factor of finesse for membranes with polarizability $\zeta(\lambda)=(n^2-1)/(2n)\sin(2\pi nl/\lambda)$, $\phi=2\pi d/\lambda$ and $\psi$ a phase factor which depends on $n$, $l$ and $\lambda$ and whose exact expression we do not reproduce here.

In addition, to take into account the small, but not completely negligible, variation of the membrane refractive index in the range considered, independent ellipsometry measurements were performed on samples originating from the same fabrication batch. The results were consistent with an index variation following a Cauchy law of the form $n(\lambda)=n_0+n_1/\lambda^2$, with $n_0=1.966\pm 0.001$ and $n_1=(1.763\pm 0.032)\times 10^{4}$ nm$^{2}$, in good agreement with what is expected for stochiometric silicon nitride. The solid line in Fig.~\ref{fig:9um}a shows the result of a fit to the theoretical model assuming the previously determined refractive index variation and with $l$ and $d$ left as free parameters. The membrane thickness and spacing resulting from this fit are $l=92.3\pm0.1$ nm and $d=8570.3\pm0.1$ nm, respectively. The extracted value for the spacing is consistent with the results of profilometer and AFM measurements performed on chips from the same batch, although the latter measurements are typically about three orders of magnitude less precise.

\begin{figure}[t]
\centering
\includegraphics[width=\linewidth]{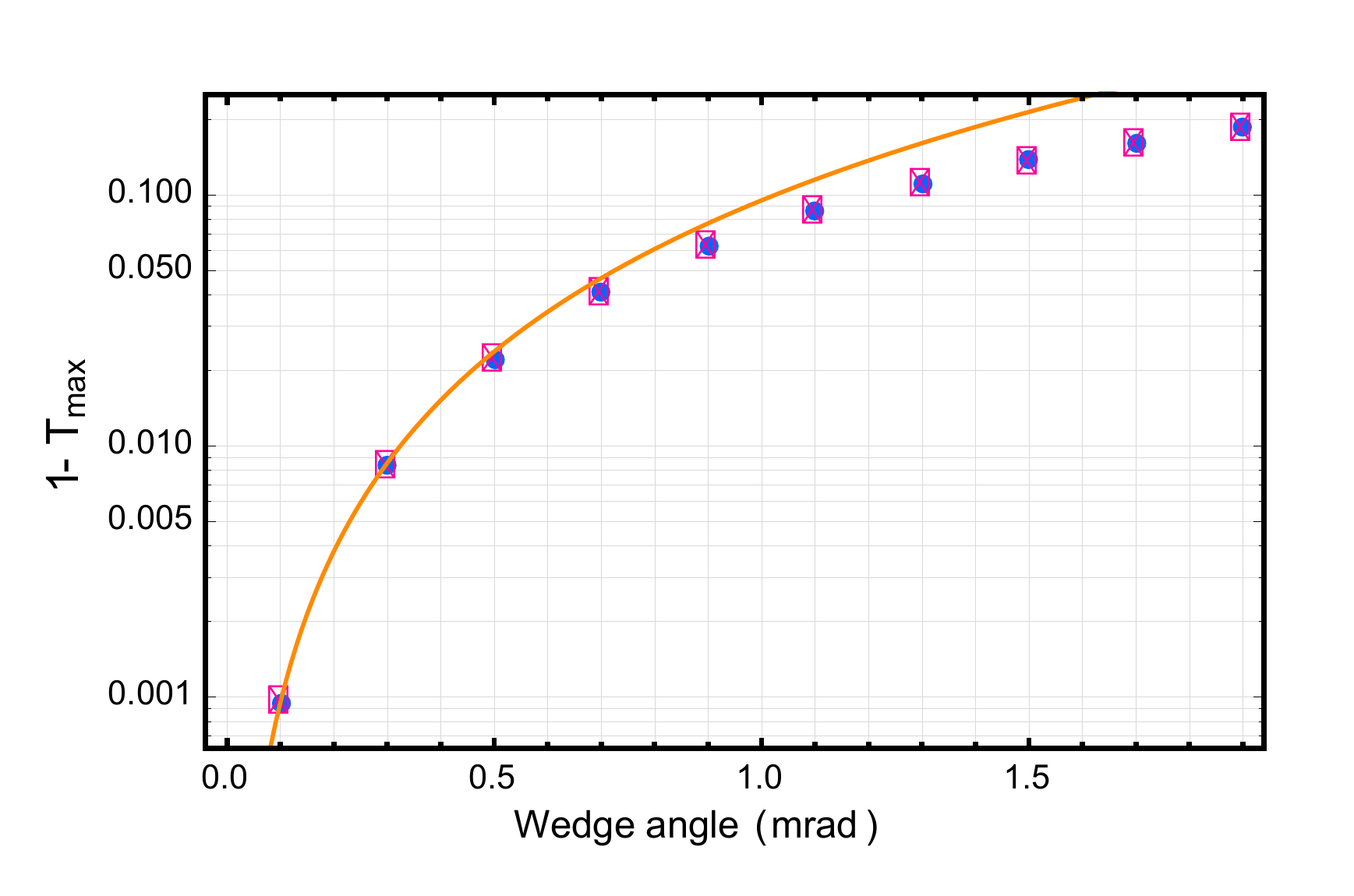}
\caption{Theoretical peak transmission deviation from unity, $1-T_{\textrm{max}}$ as a function of the wedge angle $\epsilon$ between the membranes for $w_0=50$ $\mu$m and for 35\%-reflectivity membranes ($\lambda\sim 900$ nm) with $n=1.99$, $d=8.57$ $\mu$m ({\color{red}$\square$}) and $d=200$ $\mu$m ({\color{blue}$\bullet$}). The yellow line shows the approximate expression of Eq.~(\ref{eq:Tmax}).} 
\label{fig:tiltedFP}
\end{figure}

Due to the limited resolution of the fiber spectrometer ($\sim$~0.3~nm), however, the broadband illumination method does not allow for a precise determination of the peak transmission of the array, which is a relevant quantity, e.g., for the insertion of such arrays in optical resonators~\cite{Xuereb2012}. To accurately determine the peak transmission, monochromatic light from a tunable external cavity diode laser is focused onto the array (waist $w_0\sim$~50~$\mu$m) and self-referenced measurements of the transmission are performed in the wavelength range available for the laser. The normalized transmission spectrum is shown in Fig.~\ref{fig:9um}b. Accurate measurements of the peak transmission around 899 mn yield a value of $99.7~\pm~0.1$~\%. Since ellipsometry measurements suggest a fair degree of uniformity for membranes from the same batch in terms of refractive index and thickness and low absorption levels which cannot account for this non-unity transmission, AFM and profilometer measurements of the spacer height and roughness on similar chips were performed. A conservative estimate for the intrinsic wedge due to spacer height variations was found to be of the order of 0.1 mrad. Figure~\ref{fig:tiltedFP} shows the results of numerical simulations of the peak transmission which take into account the Gaussian nature of the incident beam and include a potential wedge between the membranes~\cite{Lee2002}. For wedge angles $\epsilon<1$ mrad the numerical results agree well with the approximate expression at lowest order in $\epsilon$ \begin{equation}T_{\textrm{max}}\simeq 1-(m\pi w_0/\lambda)^2\epsilon^2\,,\label{eq:Tmax}\end{equation} obtained when one neglects the Gaussian beam curvature and considers low reflectivity dielectrics. The observed deviation from unity transmission is thus consistent with an imperfect parallelism of the membranes, corresponding to a wedge angle of $\sim0.2$ mrad, either intrinsic or imposed during assembly.

\begin{figure}[t]
\centering
\includegraphics[width=\linewidth]{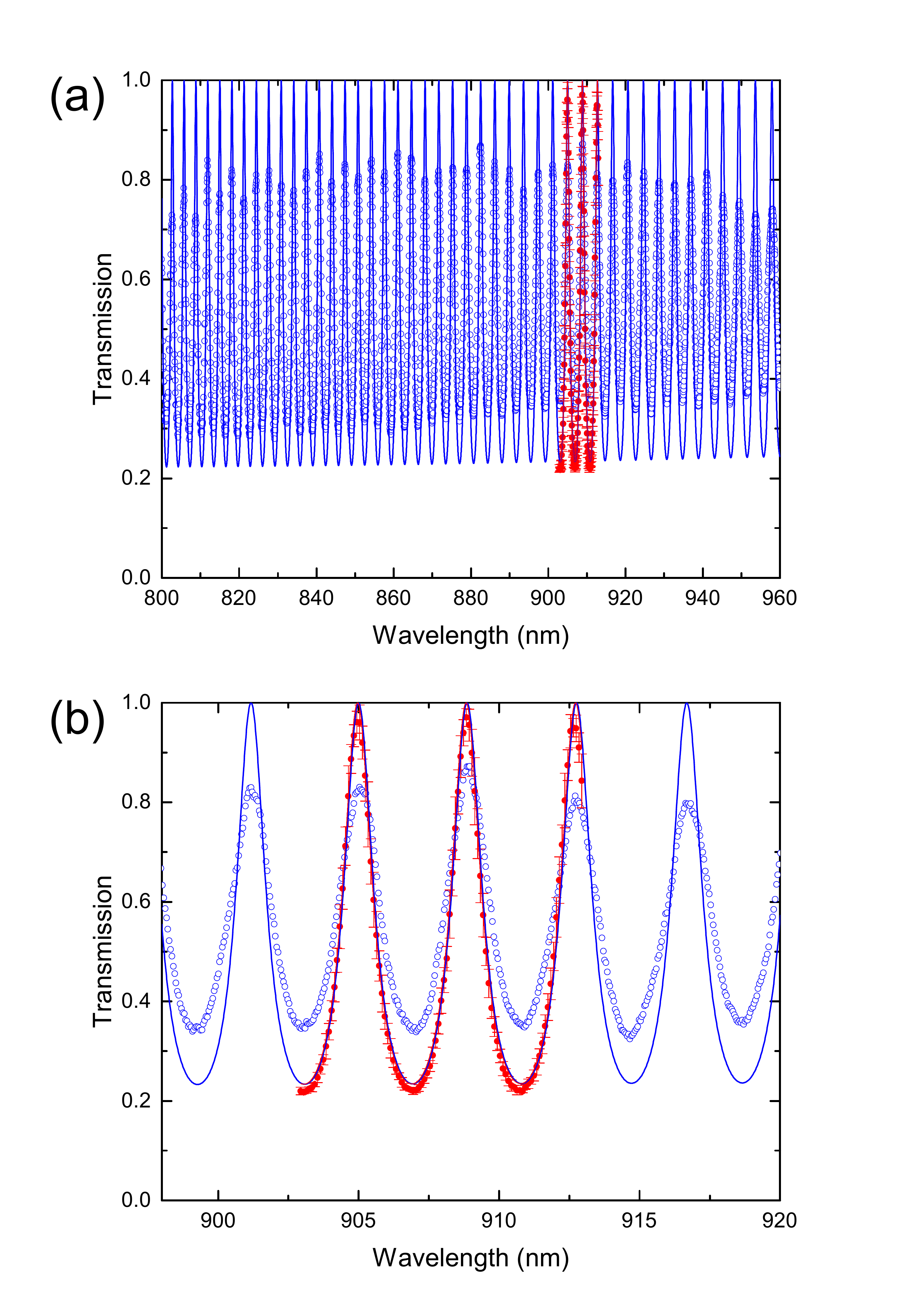}
\caption{(a) Transmission spectrum in the range 800-960 nm of a 100 $\mu$m-spacing array measured with broadband ({\color{blue}$\circ$}) and monochromatic ({\color{red}$\bullet$}) illuminations. The solid lines represent the results of fits of the data with the theoretical model. (b) Same data over the range 898-920 nm.}
\label{fig:100um}
\end{figure}

Similar measurements were performed on arrays assembled as shown in Fig.~\ref{fig:fig1}b with an intermembrane spacing dictated by the silicon frame thickness of 100 or 200 $\mu$m. The obtained transmission spectra under broadband and monochromatic illuminations for a 100 $\mu$m-spacing array are shown over the range 800-970 nm in Fig.~\ref{fig:100um}a and 898-920 nm in Fig.~\ref{fig:100um}b. Even though a $\sim 0.1$ nm-resolution fiber spectrometer is used in the broadband illumination measurement to better resolve the finer structure, the contrast of the observed interferences is substantially reduced as compared to that observed under monochromatic illumination (Fig.~\ref{fig:100um}b). The interference patterns obtained in both measurements nicely overlap though. The results of fits of the broadband and monochromatic illumination data to the theoretical model leaving $l$ and $d$ as free parameters and using an ellipsometry-determined Cauchy law refractive index variation with $n_0=1.971\pm 0.004$ and $n_1=(1.589\pm0.033)\times 10^{4}$ nm$^{2}$, yield $l=100.0\pm0.3$ nm and $d=106\,359\pm4$ nm for the broadband illumination data and $l=100.0\pm0.4$ nm and $d=106\,355\pm5$ nm, respectively. Remarkably, owing to the large number of observable interference periods, the intermembrane spacing can be quite precisely determined using the broadband illumination spectrum, even in presence of a reduced contrast. The peak transmission for this sample around 900 nm was $\sim$ 97\%, limited again by imperfect parallelism during assembly. This level was consistent with profilometer-determined frame thickness wedge angles of the order of 0.5 mrad. Similar results were obtained with commercial samples having a frame thickness of 200 $\mu$m.

\begin{figure}[htbp]
\centering
\includegraphics[width=\linewidth]{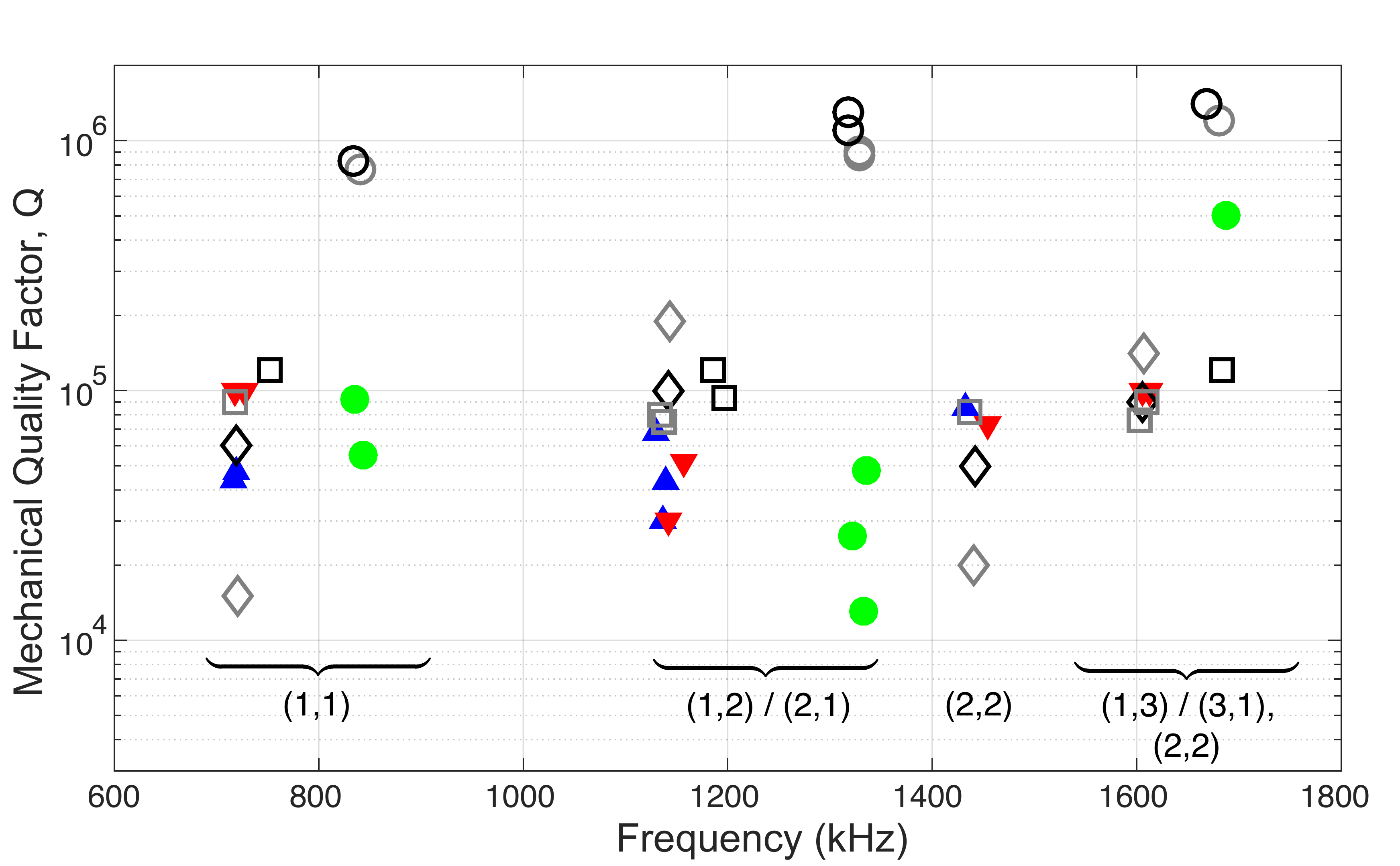}
\caption{Mechanical quality factors for the lowest frequency modes of single membranes ($\square$, {\color{defgray}$\square$}: 500 $\mu$m-thick frame without spacer, $\diamond$, {\color{defgray}$\diamond$}: 500 $\mu$m-thick frame with 8.5 $\mu$m spacer, $\circ$, {\color{defgray}$\circ$}: 200 $\mu$m thick frame) and double-membrane arrays ({\color{red}$\blacktriangledown$}, {\color{blue}$\blacktriangle$} 8.5$\mu$m-spacing, {\color{green}$\bullet$}: 200 $\mu$m-spacing). {\color{blue}$\blacktriangle$} corresponds to the array whose spectrum is shown in Fig.~\ref{fig:9um}. The corresponding square drum mode indices are indicated below the data points.}
\label{fig:mech}
\end{figure}

After the optical characterization (in air) of the samples their vibrational noise spectra were measured in a low-pressure environment ($\sim10^{-6}$ mbar) using a $\sim6$ mm-long interferometer consisting of a 50:50 beamsplitter and the sample. Light from the narrow linewidth external cavity diode laser at $\sim 900$ nm was focused onto the sample (waist $\sim60$ $\mu$m), which was resting on the frame corners, and the light transmitted from the interferometer was detected with a low-noise transimpedance photodetector. A pizeoelectric transducer inserted between the beamsplitter and the sample allowed for tuning the length of the interferometer and for performing ringdown measurements of the membranes' vibrational modes~\cite{Thompson2008} in order to establish their mechanical quality factors. The results of characteristic measurements are shown in Fig.~\ref{fig:mech}. The normal mode frequencies of the membranes used to make arrays with the 8.5 $\mu$m-inbuilt spacer were found to be very close to those of square drums with a tensile stress of $\sim 715$ MPa. Without preselection the frequency spacing between the fundamental modes of both membranes in the array was typically observed to be of a few kHz, reflecting the natural spread within the batch. While some arrays showed substantial deterioration of their mechanical quality factors after optical characterization and manipulation in an unclean environment, a few samples showed modes with Q-factors comparable to those of single membranes from the same batch ($10^5$), as can be seen from Fig.~\ref{fig:mech}. Similar measurements were performed as well with 200 $\mu$m-spacing arrays, where the average tensile stress of the individual membranes from this particular batch was higher ($\sim 945$ MPa), as well as the mechanical Q-factors ($\sim 10^6$). The fact that some arrays still displayed modes with similar Q values as those measured on single membranes from the same batch, even after extensive manipulation and characterization in air, suggests that, by minimizing manipulations after assembly, it should be possible to preserve the individual membrane mechanical properties, as is needed for future optomechanical experiments. 

In conclusion, double-membrane arrays consisting of high-tensile stress, 100-nm thick silicon nitride suspended films with separations ranging from 8.5 to 200 $\mu$m, were assembled and characterized both optically and mechanically. The arrays were shown to have good parallelism and optical properties as well as promising mechanical quality. These results, together with the  characterization techniques which are amenable to arrays with more than two membranes, represent the first experimental steps towards studies of collective optomechanics with nanomembrane arrays.

We acknowledge support from the Danish National Council for Independent research (Sapere Aude program), Villumfonden, Carlsbergfondet and the European Commission (ITN ``CCQED''). The authors thank Gunhild R. Thorsen for helping with array assembly and optical characterization.

\bibliography{optical_membrane_arrays_bib.bib}

\begin{thebibliography}{10}
\newcommand{\enquote}[1]{``#1''}

\bibitem{Aspelmeyer2014}
M.~Aspelmeyer, T.~J. Kippenberg, and F.~Marquardt, Rev. Mod. Phys. \textbf{86},
  1391 (2014).

\bibitem{Bhattacharya2008multiple}
M.~Bhattacharya and P.~Meystre, Phys. Rev. A \textbf{78}, 041801 (2008).

\bibitem{Hartmann2008}
M.~J. Hartmann and M.~B. Plenio, Phys. Rev. Lett. \textbf{101}, 200503 (2008).

\bibitem{Ludwig2012}
M.~Ludwig, A.~H. Safavi-Naeini, O.~Painter, and F.~Marquardt, Phys. Rev. Lett.
  \textbf{109}, 063601 (2012).

\bibitem{Xuereb2012}
A.~Xuereb, C.~Genes, and A.~Dantan, Phys. Rev. Lett. \textbf{109}, 223601
  (2012).

\bibitem{Stannigel2012}
K.~Stannigel, P.~Komar, S.~J.~M. Habraken, S.~D. Bennett, M.~D. Lukin,
  P.~Zoller, and P.~Rabl, Phys. Rev. Lett. \textbf{109}, 013603 (2012).

\bibitem{Seok2012}
H.~Seok, L.~F. Buchmann, S.~Singh, and P.~Meystre, Phys. Rev. A \textbf{86},
  063829 (2012).

\bibitem{Heinrich2011}
G.~Heinrich, M.~Ludwig, J.~Qian, B.~Kubala, and F.~Marquardt, Phys. Rev. Lett.
  \textbf{107}, 043603 (2011).

\bibitem{Ludwig2013}
M.~Ludwig and F.~Marquardt, Phys. Rev. Lett. \textbf{111}, 073603 (2013).

\bibitem{Zhang2012}
M.~Zhang, G.~S. Wiederhecker, S.~Manipatruni, A.~Barnard, P.~McEuen, and
  M.~Lipson, Phys. Rev. Lett. \textbf{109}, 233906 (2012).

\bibitem{Bagheri2013}
M.~Bagheri, M.~Poot, L.~Fan, F.~Marquardt, and H.~X. Tang, Phys. Rev. Lett.
  \textbf{111}, 213902 (2013).

\bibitem{Xuereb2014}
A.~Xuereb, C.~Genes, G.~Pupillo, M.~Paternostro, and A.~Dantan, Phys. Rev.
  Lett. \textbf{112}, 133604 (2014).

\bibitem{Xuereb2015}
A.~Xuereb, A.~Imparato, and A.~Dantan, New J. Phys. \textbf{17}, 055013 (2015).

\bibitem{Chesi2014}
S.~Chesi, Y.-D. Wang, and J.~Twamley, Sci. Rep. \textbf{5}, 7816 (2014).

\bibitem{Schmidt2015}
M.~Schmidt, V.~Peano, and F.~Marquardt, New J. Phys. \textbf{17}, 023025
  (2015).

\bibitem{Massel2012}
F.~Massel, S.~U. Cho, J.-M. Pirkkalainen, P.~J. Hakonen, T.~T. Heikkil\"a, and
  M.~A. Sillanp\"a\"a, Nat. Commun. \textbf{3}, 987 (2012).

\bibitem{Ockeloen2016}
C.~F. Ockeloen-Korppi, E.~Damsk\"agg, J.-M. Pirkkalainen, A.~A. Clerk, M.~J.
  Woolley, and M.~A. Sillanp\"a\"a, Phys. Rev. Lett. \textbf{117}, 140401
  (2016).

\bibitem{Dong2012}
C.~Dong, V.~Fiore, M.~C. Kuzyk, and H.~Wang, Science \textbf{338}, 1609 (2012).

\bibitem{Zhang2015}
M.~Zhang, S.~Shah, J.~Cardenas, and M.~Lipson, Phys. Rev. Lett. \textbf{115},
  163902 (2015).

\bibitem{Gilsantos2016}
E.~Gil-Santos, M.~Labousse, C.~Baker, A.~Goetschy, W.~Hease, C.~Gomez,
  A.~Lemaitre, G.~Leo, C.~Ciuti, and I.~Favero, arxiv:1609.09712  (2016).

\bibitem{Fang2016}
K.~Fang, M.~H. Matheny, X.~Luan, and O.~Painter, Nature Photonics \textbf{7},
  489 (2016).

\bibitem{Botter2013}
T.~Botter, D.~W.~C. Brooks, S.~Schreppler, N.~Brahms, and D.~M. Stamper-Kurn,
  Phys. Rev. Lett. \textbf{110}, 153001 (2013).

\bibitem{Spethmann2016}
N.~Spethmann, J.~Kohler, S.~Schreppler, L.~Buchmann, and D.~M. Stamper-Kurn,
  Nat. Phys. \textbf{12}, 27 (2016).

\bibitem{Thompson2008}
J.~D. Thompson, B.~M. Zwickl, A.~M. Jayich, F.~Marquardt, S.~M. Girvin, and
  J.~G.~E. Harris, Nature \textbf{452}, 72 (2008).

\bibitem{Wilson2009}
D.~J. Wilson, C.~A. Regal, S.~B. Papp, and H.~J. Kimble, Phys. Rev. Lett.
  \textbf{103}, 207204 (2009).

\bibitem{Karuza2012}
M.~Karuza, C.~Molinelli, M.~Galassi, C.~Biancofiore, R.~Natali, P.~Tombesi,
  G.~D. Giuseppe, and D.~Vitali, New J. Phys. \textbf{14}, 095015 (2012).

\bibitem{Kemiktarak2012NJP}
U.~Kemiktarak, M.~Durand, M.~Metcalfe, and J.~Lawall, New J. Phys. \textbf{14},
  125010 (2012).

\bibitem{Purdy2013}
T.~P. Purdy, R.~W. Peterson, and C.~A. Regal, Science \textbf{339}, 801 (2013).

\bibitem{Kemiktarak2014}
U.~Kemiktarak, M.~Durand, M.~Metcalfe, and J.~Lawall, Phys. Rev. Lett.
  \textbf{113}, 030802 (2014).

\bibitem{Shkarin2014}
A.~B. Shkarin, N.~E. Flowers-Jacobs, S.~W. Hoch, A.~D. Kashkanova, C.~Deutsch,
  J.~Reichel, and J.~G.~E. Harris, Phys. Rev. Lett. \textbf{112}, 013602
  (2014).

\bibitem{Lee2015}
D.~Lee, M.~Underwood, D.~Mason, A.~B. Shkarin, S.~W. Hoch, and J.~G.~E. Harris,
  Nat. Commun. \textbf{6} (2015).

\bibitem{Xu2016}
m.~D. J.~L. Xu, H. and J.~G.~E. Harris, Nature \textbf{80}, 537 (2016).

\bibitem{Noguchi2016}
A.~Noguchi, R.~Yamazaki, M.~Ataka, H.~Fujita, Y.~Tabuchi, T.~Ishikawa,
  K.~Usami, and Y.~Nakamura, arxiv:1602.01554  (2016).

\bibitem{Nielsen2016}
W.~H.~P. Nielsen, Y.~Tsaturyan, C.~B. M{\o}ller, E.~S. Polzik, and
  A.~Schliesser, arxiv:1605.06541  (2016).

\bibitem{Bagci2014}
T.~Bagci, A.~Simonsen, S.~Schmid, L.~G. Villanueva, E.~Zeuthen, J.~Appel, J.~M.
  Taylor, A.~Sorensen, K.~Usami, A.~Schliesser, and E.~S. Polzik, Nature
  \textbf{507}, 81 (2014).

\bibitem{Andrews2014}
R.~W. Andrews, R.~W. Peterson, T.~P. Purdy, K.~Cicak, R.~W. Simmonds, C.~A.
  Regal, and K.~W. Lehnert, Nat. Phys. \textbf{10}, 321 (2014).

\bibitem{Fink2016}
J.~M. Fink, M.~Kalaee, A.~Pitanti, R.~Norte, L.~Heinzle, M.~Davanço,
  K.~Srinivasan, and O.~Painter, Nature Communications \textbf{7}, 12396
  (2016).

\bibitem{Xuereb2013}
A.~Xuereb, C.~Genes, and A.~Dantan, Phys. Rev. A \textbf{88}, 053803 (2013).

\bibitem{Tignone2014}
E.~Tignone, G.~Pupillo, and C.~Genes, Phys. Rev. A \textbf{90}, 053831 (2014).

\bibitem{Nair2016}
B.~Nair, A.~Xuereb, and A.~Dantan, Phys. Rev. A \textbf{94}, 053812 (2016).

\bibitem{Lee2002}
J.~Y. Lee, J.~W. Hahn, and H.-W. Lee, J. Opt. Soc. Am. A \textbf{19}, 973
  (2002).

\end{thebibliography}

\end{document}